\documentclass[a4paper]{jpconf}
\usepackage{graphicx}
\usepackage[hidelinks]{hyperref}
\usepackage{siunitx}
\usepackage{cite}
\newcommand{\msr}{$\mu$SR} 
 
\newcommand{\lem}{LE-$\mu$SR}

\begin{document}
\title{Thin film and surface preparation chamber for the low energy muons spectrometer}

\author{Hanna Teuschl$^{1,2}$, Angelo Di Bernardo$^3$, Leandro M. O. Lourenço$^4$, Thomas Prokscha$^1$, Ricardo B. L. Vieira$^5$ and Zaher Salman$^1$}
\address{$^1$ Paul Scherrer Institut, Laboratory for Muon Spin Spectroscopy, CH-5232 Villigen PSI, Switzerland}
\address{$^2$ Institute of Physics, Montanuniversität Leoben, Franz Josef Str. 18, 8700 Leoben, Austria}
\address{$^3$ Department of Physics, University of Konstanz, 78457, Konstanz, Germany}
\address{$^4$ LAQV–REQUIMTE, Department of Chemistry, University of Aveiro, 3810–193 Aveiro, Portugal}

\ead{zaher.salman@psi.ch}

\begin{abstract}
We have designed and constructed a thin film preparation chamber with base pressure of $<2 \times 10^{-9}$~mbar. Currently, the chamber is equipped with two large area evaporators (a molecular evaporator and an electron-beam evaporator), an ion sputtering gun, a thickness monitor and a substrate heater. It is designed such that it can handle large area thin film samples with a future possibility to transfer them in vacuum directly to the low energy muons (LEM) spectrometer or to other advanced characterization facilities in the Quantum Matter and Materials Center (QMMC) which will be constructed in 2024. Initial commissioning of the chamber resulted in high quality, large area and uniform molecular films of CuPc and TbPc$_2$ on various substrate materials. We present first results from low energy \msr\ (\lem) measurements on these films.
\end{abstract}

\section{Introduction}
Multilayers of molecular/inorganic quantum materials present a new approach to designing systems with novel properties which cannot be achieved using alternative methods. The low cost of molecular materials, the tunability of their physical properties and the long spin coherence times make them an ideal multifunctional platform for quantum computation and spintronic applications \cite{Warner2013N,Cornia2011CSR}. However, their deposition in thin film layers and their hybrid interfaces with different materials, generally present a non-trivial change to their properties which can be crucial for any future application \cite{Hofmann2012AN,Wackerlin2016AM,Serrano2020NM}. Such systems have been the focus of intense research in recent years, in particular with regards to the properties of one side of the hybrid interface, i.e., the molecular side \cite{Hofmann2012AN,Wackerlin2016AM,Serrano2020NM,Kiefl2016AN,Cornia2011CSR}. This leaves the potential of discovery of new phenomena in the inorganic side largely unexplored. In part, the limited access of the available experimental techniques to buried interfaces, makes it extremely difficult to investigate them.

The low energy \msr\ (\lem) techniques provides a perfectly suited method to investigate such systems in a depth resolved manner. It enables a systematic investigation of the variation in the magnetic and electronic properties across the molecular/inorganic interface using the same probe. However, it comes with its own requirements and limitations concerning the large sample area, thickness and homogeneity of the layers. Therefore, in order to be able to investigate such interfaces with \lem, we constructed a sample preparation chamber, named Leyla, specifically designed to fabricate and handle large area samples. In addition to the thin film deposition capabilities, the chamber will be equipped with surface preparation and characterization capabilities. Moreover, a load-lock (LL) system attached to Leyla will enable future in-vacuum transport of samples to the low energy muons (LEM) spectrometer \cite{Prokscha2008NIPA} as well as other advanced characterization tools across the Paul Scherrer Institute (PSI).

In this paper, we focus on thin film deposition of Copper(II) phthalocyanine (CuPc) and the single molecule magnet (SMM) Terbium(III) phthalocyanine (TbPc$_2$), starting from bulk powders, on various substrate materials. We demonstrate that we can fabricate highly uniform films on a large surface area (\SI{>2}{\cm\squared}) with thickness variation of less than \SI{\sim 10}{\percent}. Furthermore, we show that the deposited molecular magnetic layers display the expected qualitative magnetic properties without obvious degradation.

\section{Experimental}
\begin{figure}[ht]
    \centering
    \includegraphics[width=10cm]{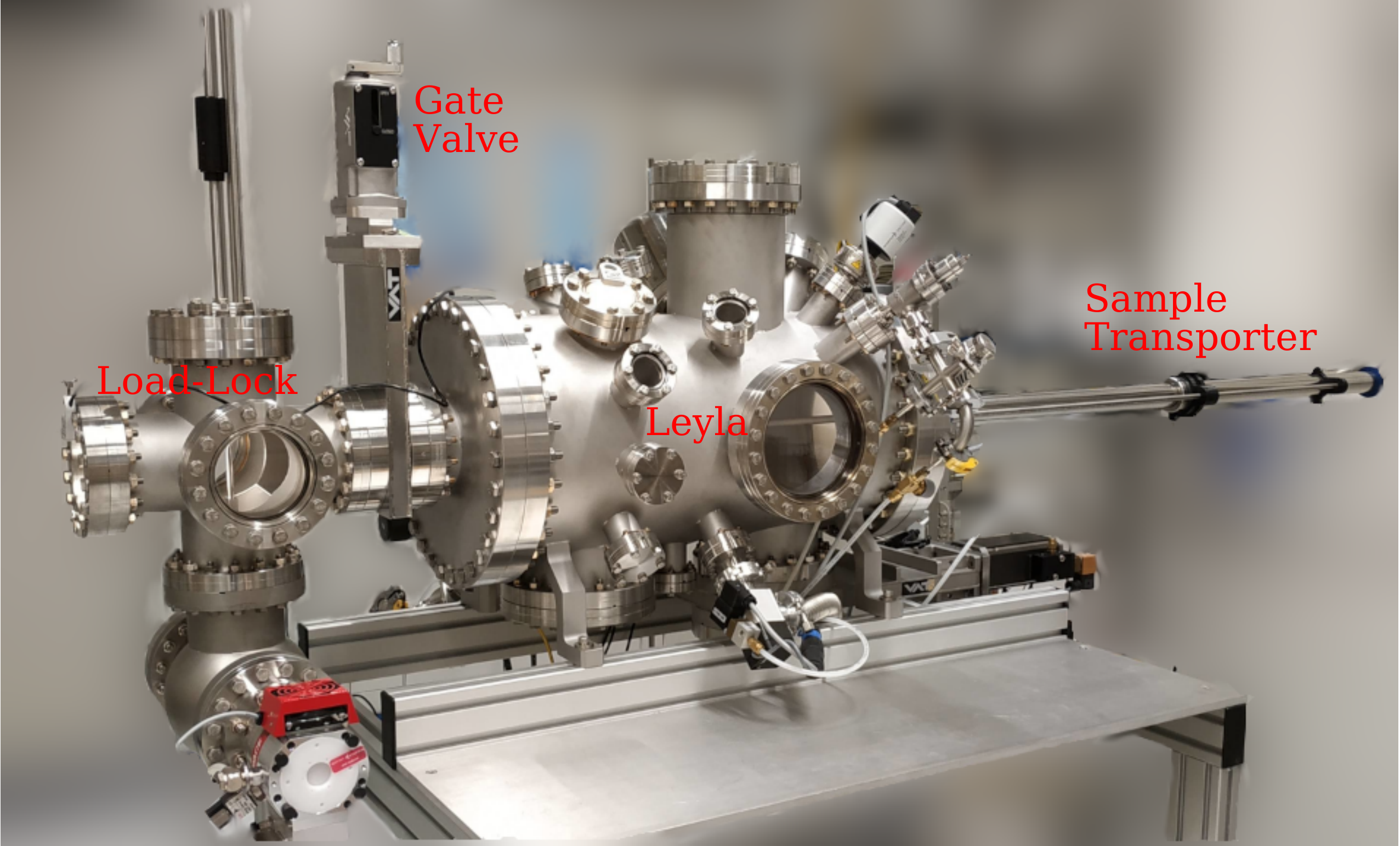}
    \caption{A photograph of the UHV preparation chamber. The load-lock on the left is separated from the main UHV chamber by a manual gate valve. The sample transporter (Ferrovac AG) is long enough to transport samples from the load-lock into the main chamber for preparation. The load-lock and Leyla are equipped with their independent pumping stations.}
    \label{Leyla}
\end{figure}

The ultra high vacuum (UHV) preparation chamber, Leyla, was designed and fabricated at PSI (see Fig.~\ref{Leyla}).
The base pressure in Leyla can go down to $\sim 4\times10^{-10}$~mbar using a single large turbo pump. The LL with its independent pumping station is connected to Leyla \emph{via} a manual gate-valve. The LL can host up to three samples which can be then inserted without breaking vacuum into Leyla. It is also envisaged that these samples can be taken out of the LL to be transported in vacuum to the LEM spectrometer or other characterization systems. Leyla has multiple ports (DN40 CF, DN63 CF and DN150 CF) that can host additional evaporators or other compatible equipment to be used for sample preparation, characterization, or manipulation. Currently, Leyla is equipped with a 3-cell molecular evaporator (Kentax TCE-CS 5x), a large area electron-beam evaporator (FOCUS FEM 4), a large cross section ion source for gas sputtering (SPECS IQP 10/63), a Pyrolytic Boron Nitride ceramic heater (tectra HTR-1002) and a quartz balance thickness monitor (PREVAC TM14).

The films studied here were deposited using the Kentax evaporator, which uses tantalum wires to heat three crucibles independently. This evaporator is perfectly suited for preparation of molecular thin layers due to their relatively low sublimation temperature. Three different thin films samples were studied here, (I) a \SI{\sim 120}{\nm} CuPc/Al$_2$O$_3$, (II) a \SI{\sim 100}{\nm} TbPc$_2$/MgO and (III)  a \SI{\sim 130}{\nm} TbPc$_2$/\SI{65}{\nm} Nb/SiO$_2$ terminated Si substrate. 

Prior to deposition, the high purity CuPc powders (Sigma-Aldrich) were baked in vacuum ($\sim \times10^{-6}$~mbar) to remove any adsorbed gasses and contaminants. Sample I was then deposited with the crucible at \SI{325}{\celsius} and at a distance of \SI{9}{\cm} from the substrate for a duration of \SI{\sim 4}{\hour}. According to our \emph{ex situ} thickness calibration measurements using a profilometer (Veeco Dektak 8), this procedure results in a highly uniform coverage of the Al$_2$O$_3$ substrate with CuPc layer thickness of \SI{\sim 120}{\nm}.

For the TbPc$_2$ powders, which where synthesized following Ref.~\cite{Deng2015PS}, we optimized a lengthy purification procedure to remove the crystallization solvents and any organic contaminants. For this, the crucible was filled with TbPc$_2$ powders (as prepared) and heated gradually in vacuum ($10^{-5}--10^{-6}$~mbar), from \SI{120}{\celsius} up to \SI{400}{\celsius} in steps of \SI{25}{\celsius}, waiting for \SIrange{1}{2}{\hour} after each step (at \SI{325}{\celsius} waiting \SI{7}{\hour}). The evaporator was then inserted into the Leyla's UHV to start the film deposition. Sample II was grown with the crucible at \SI{400}{\celsius} and at a distance of \SI{7}{\cm} from the substrate for a duration of \SI{7}{\hour}. Sample III, on the other hand, was prepared with the crucible at the same temperature, \SI{400}{\celsius}, but at a distance of \SI{8}{\cm} from the substrate for a duration of \SI{9}{\hour}. With these deposition conditions we aimed for a TbPc$_2$ layer thickness of \SIrange{120}{150}{\nm} but with a higher thickness uniformity.

\section{Results}
\subsection{CuPc/Al$_2$O$_3$}
\begin{figure}[ht]
    \centering
    \includegraphics[height=6cm]{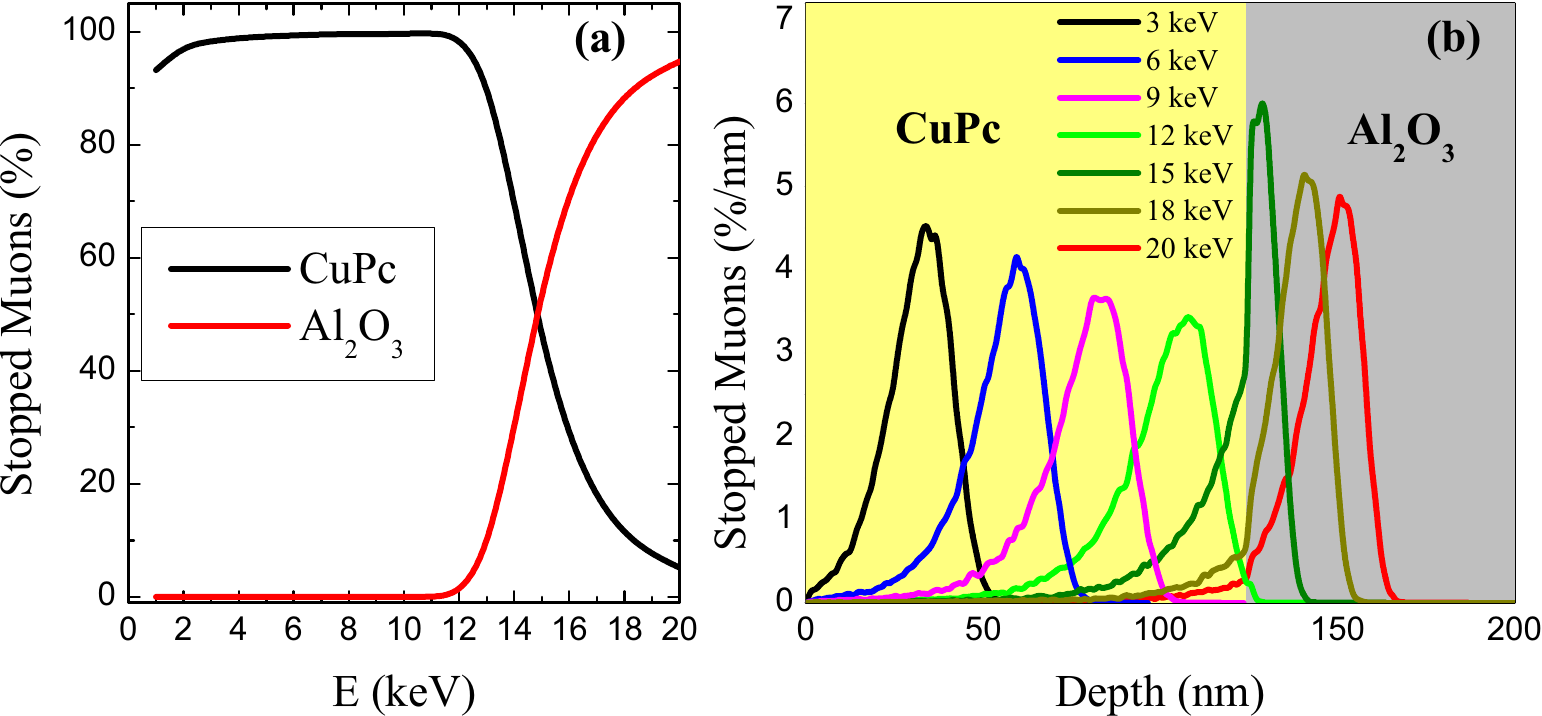}
    \caption{Results of TRIM.SP simulations of muons implanted in a \SI{125}{\nm} film of CuPc/Al$_2$O$_3$. (a) The fraction of stopped muons as a function of $E$ in the CuPc layer (black line) and the Al$_2$O$_3$ substrate (red line). (b) The muon stopping profiles as a function of depth for various $E$.}
    \label{ProfilesCuPc}
\end{figure}

The CuPc molecule has an unpaired electron at the Cu ion leading to a paramagnetic behaviour down to cryogenic temperatures \cite{PirotoDuarte2006PRB}. To estimate the thickness of the film and its uniformity we performed \lem\ measurements in a weak transverse field (TF) of \SI{10}{mT} at $T=290\,\si{\K}$. At this temperature, muons stopping in the paramagnetic CuPc are expected to give a large precession signal, while those stopping in the Al$_2$O$_3$ form predominantly muonium (Mu) which does not contribute to the asymmetry precessing at the Larmor frequency \cite{Krieger2017PRB,Vilao2021PRB}. Therefore, by implanting the muons with increasing energy, $E$, we expect to observe a decrease in the precessing asymmetry amplitude as more muons reach the Al$_2$O$_3$ substrate. In Fig.~\ref{ProfilesCuPc} we present results from TRIM.SP simulations \cite{Eckstein1991,Morenzoni2002NIMPRSB} of muons implanted in a \SI{125}{\nm} film of CuPc on Al$_2$O$_3$, which show that a fraction of the implanted muons will stop in the substrate for $E$ higher than \SI{\sim 12}{\keV}.

\begin{figure}[hb]
    \centering
    \includegraphics[height=6cm]{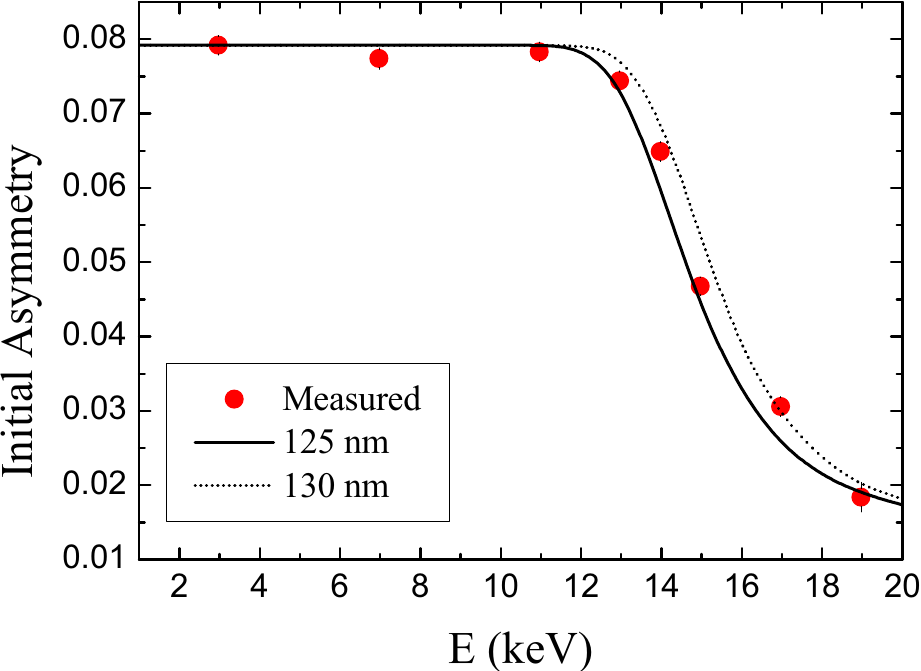}
    \caption{The initial asymmetry measured in the CuPc/Al$_2$O$_3$ film, obtained from the fits of the \lem\ data. The solid and dotted lines represent the calculated value of the initial asymmetry based on TRIM.SP simulations for a \SI{125}{\nm} and \SI{130}{\nm} thick CuPc films, respectively.}
    \label{CuPcTrim}
\end{figure}

The \lem\ measurements in TF were fit (using musrfit \cite{Suter2012PP}) to an exponentially damped
oscillation, which exhibits a clear variation as a function of
$E$. The initial asymmetry obtained from the fits, shown in
Fig.~\ref{CuPcTrim}, is almost constant at low $E$ and starts
decreasing for $E>12\,\si{\keV}$. 
This is a clear indication that at these high energies the muons reach
the Al$_2$O$_3$ substrate. In order to more accurately estimate the
thickness of the CuPc layer we ran TRIM.SP simulations for different
CuPc layer thicknesses and calculated the weighted average of the asymmetry
contribution from CuPc and Al$_2$O$_3$. The curves calculated for CuPc
thickness of \SI{125}{\nm} (solid line) and \SI{130}{\nm} (dotted line) are shown in
Fig.~\ref{CuPcTrim}, and clearly capture the measured $E$ dependence,
giving an estimated CuPc layer thickness of \SIrange{125}{130}{\nm} in
agreement with our calibration measurements using the profilometer. 
 
Turning to the temperature dependence of the \lem\ data, we note
first that bulk \msr\ measurements on powder CuPc show that the muon
spin relaxes following a two exponential relaxation function, with
two relaxation rates that vary by about two orders of magnitude
\cite{PirotoDuarte2006PRB}. Moreover, it was found that the relaxation
rates increase gradually with decreasing temperature. The fast
relaxing component is on the order of \SI{\sim 20}{\per\us},
i.e. too high to be measured using \lem\ due to the limited time
resolution of the spectrometer. Therefore, our fits reflect only the
behaviour of the slow relaxing component, and hence give a small oscillating asymmetry amplitude (0.08) from the CuPc layer in the energy scan
(Fig.~\ref{CuPcTrim}).

\begin{figure}[ht]
    \centering
    \includegraphics[width=8cm]{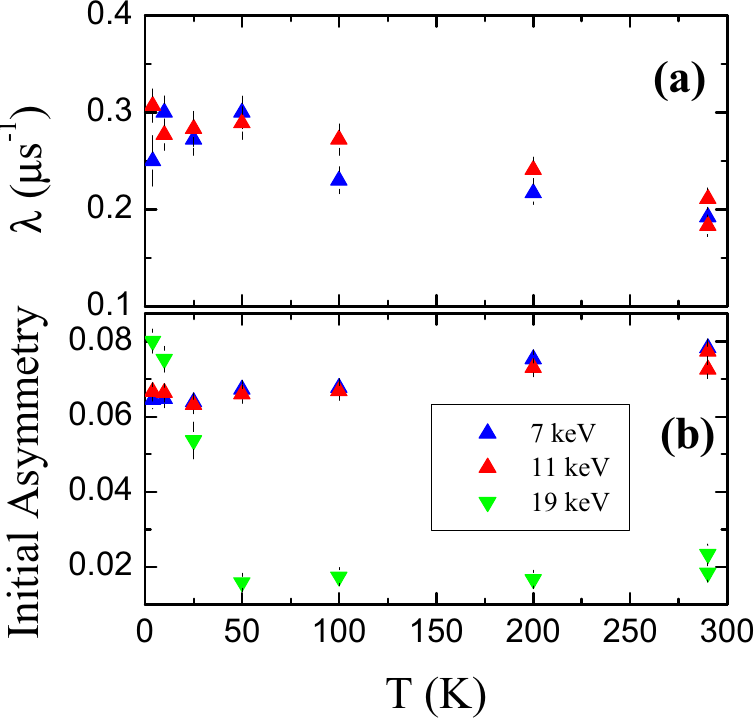}
    \caption{The temperature dependence of the (a) damping rate and
      (b) initial asymmetry measured in a TF of \SI{10}{mT} and at different $E$ in the CuPc/Al$_2$O$_3$ film.}
    \label{CuPcTScan}
\end{figure}

We performed \lem\ measurements in TF of \SI{10}{mT} as a function of temperature at three implantation energies; \SIlist{7;11;19}{\keV}. Results from fitting the data are shown in Fig.~\ref{CuPcTScan}.
The initial asymmetry of the slow relaxing component measured at $E=7$
and \SI{11}{\keV} exhibits a small decrease as a function of temperature, similar to what was observed in bulk samples \cite{PirotoDuarte2006PRB}. The damping rate measured at these
energies has a similar magnitude and temperature dependence to that observed in bulk, i.e., a small increase with decreasing temperature. Note that
we do not observe a depth dependence in the CuPc layer itself indicating that its magnetic properties are depth independent. The temperature dependence of the initial asymmetry measurements at
$E=19\,\si{\keV}$, where almost all the implanted muons stop in the sapphire
substrate (Fig.~\ref{CuPcTrim}), follows what we expect from this substrate \cite{Krieger2017PRB,Vilao2021PRB}, confirming yet again the agreement between TRIM.SP simulations and the experimental measurements.

\subsection{TbPc$_2$/MgO and TbPc$_2$/Nb}
The TbPc$_2$ molecules have a double-decker structure with a Tb$^{3+}$ ion
sandwiched between two Pc molecules \cite{Ishikawa2003JACS}. This is one of the most
extensively studied SMMs due to its relatively high chemical
stability, which enables its sublimation and deposition on a wide
range of substrates. In addition, it has a large anisotropy barrier \cite{Ishikawa2003IC} resulting in a long molecular spin correlation time at relatively high temperatures \cite{Branzoli2009PRB,Hofmann2012AN}.

\begin{figure}[ht]
    \centering
    \includegraphics[height=6cm]{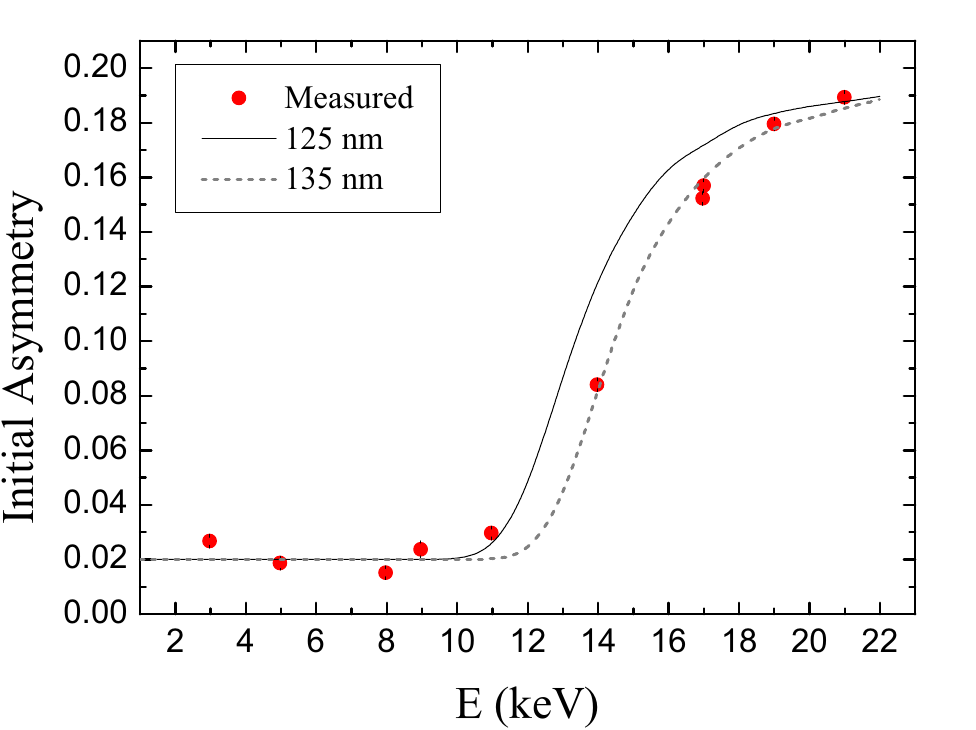}
    \caption{The initial asymmetry measured as a function of $E$ in the TbPc$_2$/Nb film. The solid and dotted lines represent the calculated values based on TRIM.SP simulations for a \SI{125}{\nm} and \SI{135}{\nm} thick TbPc$_2$ layer thickness, respectively.}
    \label{TbPc2Trim}
\end{figure}
We performed zero field (ZF) and TF \lem\ measurements on two
TbPc$_2$ thin film samples, one grown on a substrate of MgO and the other deposited on top of a thin layer of Nb metal (which was grown on a SiO$_2$ terminated Si substrate). The TF measurements at $T=290\,\si{\K}$ and \SI{5}{\K}, were used
to estimate the thickness of the TbPc$_2$ layers following the same
procedure described above for the CuPc films. For the TbPc$_2$/MgO,
the amplitude of the oscillating asymmetry decreases once the muons
reach the MgO substrate due to muonium formation. Therefore, best
contrast between the contribution from TbPc$_2$ and MgO is observed at
high temperatures where TbPc$_2$ is paramagnetic, i.e. at \SI{290}{\K}. From
these measurements we estimate the thickness of the TbPc$_2$ in this
sample to be \SIrange{95}{105}{\nm}. In contrast, muons stopping in Nb give a
fully diamagnetic signal. Therefore, best contrast between the
asymmetry contribution from TbPc$_2$ and Nb is obtained at low
temperatures (\SI{5}{\K}), where TbPc$_2$ is strongly magnetic (quasi-static)
leading to no contribution to the oscillating asymmetry. From these
measurements we obtain TbPc$_2$ layer thickness of \SIrange{125}{135}{\nm}, as
can be seen in Fig~\ref{TbPc2Trim}.

\begin{figure}[hb]
    \centering
    \includegraphics[height=6cm]{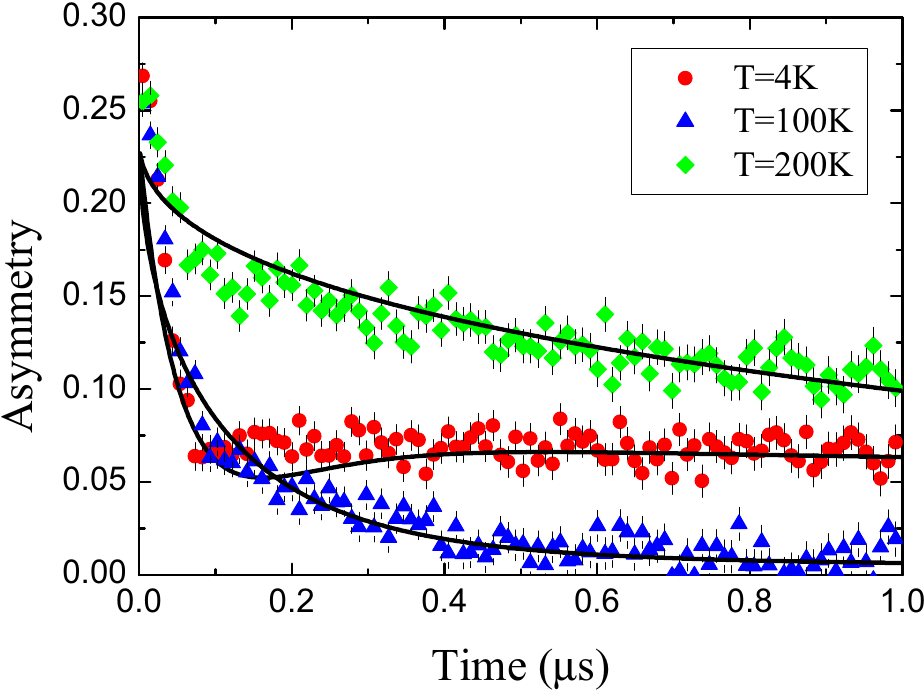}
    \caption{Typical asymmetry spectra measured in the TbPc$_2$/Nb film
      at $E=8\,\si{\keV}$ and ZF. The solid lines are fits as described in
      the text.}
    \label{TbPc2ZF}
\end{figure}

\begin{figure}[ht]
    \centering
    \includegraphics[width=8cm]{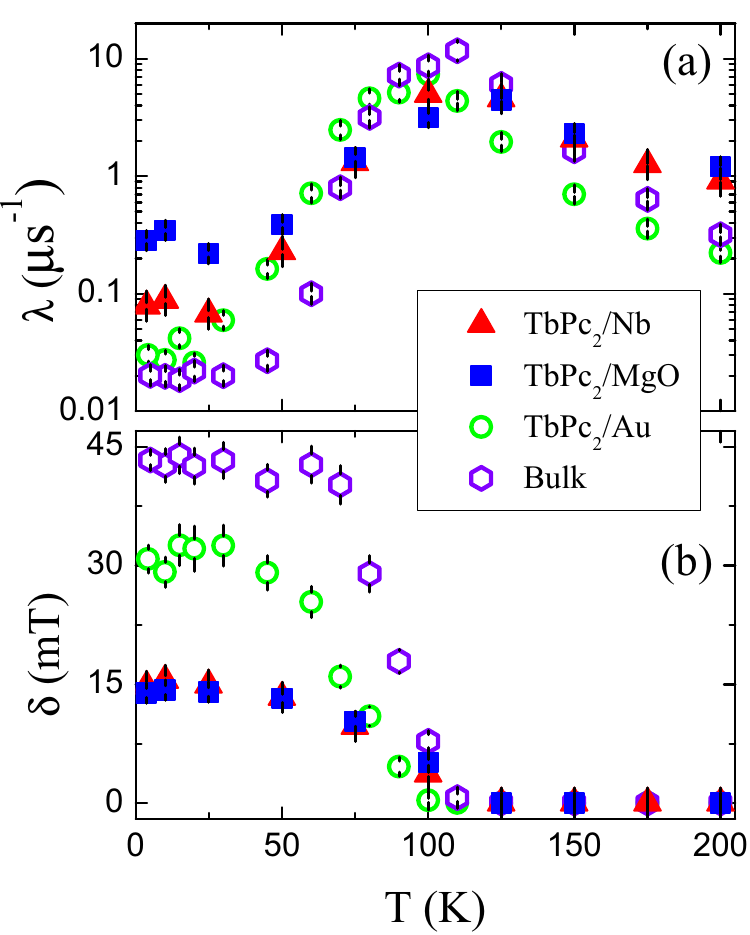}
    \caption{(a) The spin lattice relaxation rate, $\lambda$, and (b) the width of the local field distribution, $\delta$, as a function of temperature measured in the TbPc$_2$ films grown on Nb (red triangles) and on MgO (blue squares). For comparison we also plot the values measured in TbPc$_2$/Au (green circles) and bulk (purple pentagons) adapted from Ref.~\cite{Hofmann2012AN}.}
    \label{TbPc2Rlx}
\end{figure}

We now turn our focus to the ZF relaxation in the TbPc$_2$ layers as a
function of temperature. Typical asymmetry spectra measured in the
TbPc$_2$/Nb sample are shown in Fig.~\ref{TbPc2ZF}.
We find that at high temperatures, above \SI{\sim 100}{\K}, the asymmetry
follows an exponential-like relaxation, where the relaxation rate
increases with decreasing temperature. At lower temperature, the
relaxation exhibits a quasi-static behaviour, i.e. it follows a
Kubo-Toyabe-like relaxation. The same qualitative behaviour is
detected in the TbPc$_2$/MgO sample and was also seen in other films
and bulk powders of TbPc$_2$ \cite{Branzoli2009PRB,Hofmann2012AN}. To fit the data in the
whole temperature range we use the same phenomenological function used
in previous studies \cite{Hofmann2012AN,Kiefl2016AN},
\begin{equation}
  A(t)=A_0 \left[ \frac{1}{3} + \frac{2}{3} (1-\gamma_\mu \delta t
    )e^{-\gamma_\mu \delta t}\right] e^{-\sqrt{\lambda t}},
\end{equation}
where $A_0$ is the initial asymmetry, $\gamma_\mu$ is the muon
gyromagnetic ratio, $\delta$ is the width of the static field distribution
experienced by the muons and $\lambda$ is the dynamic muon spin relaxation rate (or spin lattice relaxation rate). For each TbPc$_2$ thin film sample, we fit all
temperatures at the same $E$ using a shared $A_0$ and allow only
$\delta$ and $\lambda$ to depend on the temperature. These parameters are
presented in Fig.~\ref{TbPc2Rlx}.
From these fits we can clearly see that the molecular spin fluctuations of the TbPc$_2$ slow gradually as the temperature is decreased. This is evident in the increase of $\lambda$ while $\delta=0$. Below \SI{\sim 100}{\K}, these dynamics become slow enough and comparable to the scale of the muon's lifetime (\SI{\sim 2.2}{\us}), resulting in the appearance of local static fields ($\delta>0$), accompanied by a decrease in $\lambda$. Finally, at temperatures below \SI{\sim 50}{\K}, both $\lambda$ and $\delta$ become temperature independent indicating that the low temperature molecular spin dynamics are dominated by quantum mechanical effects \cite{Branzoli2009PRB,Hofmann2012AN}.  

\section{Summary and Conclusions}
To summarize, we have successfully deposited highly homogeneous, large
area, molecular thin films suitable for \lem\ studies. In the CuPc
films we find that there are no significant changes to their magnetic
properties as a function of depth and that the \msr\ spectra and
extracted muon spin relaxation rates are consistent with those measured
in bulk powders. In contrast, we find that while the qualitative
properties of TbPc$_2$ films are similar to those measured in thin
films of TbPc$_2$ deposited on Au, the exact values of the dynamic
spin lattice relaxation rate and the width of the static fields distribution
differ. Moreover, comparing the muon spin relaxation measured in
similar films deposited on Nb metal and insulating MgO, we find that
the temperature dependence of the width of the local field
distribution is the same while the dynamic relaxation rates differs at
low temperatures, i.e. in the quasistatic regime. We also note that the difference is much larger than the statistical and systematic errors in our measurements. Therefore, the effect hints to a possible substrate effect on the TbPc$_2$ low temperature molecular spin dynamics,
which will be investigated more systematically as a function of depth in future studies.

The origin of the general difference between the properties of our
newly deposited films and those studies previously is not yet fully
understood. The smaller $\delta$ at temperatures below \SI{\sim 100}{\K} in the new films may be an indication of a
highly oriented packing of the molecules compared to the more
amorphous packing in the previously studies samples \cite{Malavolti2013JMCC,Hofmann2012AN}. Other possibilities such as the presence of some non-magnetic/paramagnetic impurities in the TbPc$_2$ layer are
highly unlikely, since it should result in a broader field distribution
from the muon's perspective. The different packing is also expected to
affect the dynamic relaxation rate over the whole temperature range which we also observe in the newly deposited films. However, thorough chemical analysis of these films is ongoing and future detailed \lem\ measurements are planned to elucidate the origin of these discrepancies. 

\ack
This work is based on experiments performed at the Swiss Muon Source
(S$\mu$S), Paul Scherrer Institute, Villigen,
Switzerland. H.T. acknowledges support by a mobility grant from Montanuniversität Leoben. 

\section*{References}
\bibliography{refs}

\end{document}